\documentclass[12pt]{article}
\usepackage{amssymb}
\usepackage{amsmath}
\usepackage{epsfig}
\usepackage{float}
\newcommand{\be}{\begin{equation}}
\newcommand{\ee}{\end{equation}}
\newcommand{\bea}{\begin{eqnarray}}
\newcommand{\eea}{\end{eqnarray}}

\begin{document}

\begin{center}

{\bf NEUTRINOLESS DOUBLE $\beta$-DECAY:THE PROBLEM OF NUCLEAR MATRIX ELEMENTS
}\footnote{
A report at the International Workshop  "Neutrino Telescopes", Venice,
February 22-25, 2005}

\end{center}
\begin{center}
S. M. Bilenky
\end{center}
\vspace{0.1cm}
\begin{center}
{\em  Joint Institute
for Nuclear Research, Dubna, R-141980, Russia\\}
\end{center}
\begin{center}
{\em SISSA via Beirut 2-4, Trieste, 34014, Italy\\}
\vspace{0.2cm}
E-mail:bilenky@he.sissa.it
\end{center}

\begin{abstract}
The calculations of nuclear matrix elements of $0\nu\beta\beta$-decay is a 
challenge for nuclear physics. We are discussing here a model independent method,
which could allow to test the calculations. The method is based on the factorization property of the nuclear matrix elements and requires observation of neutrinoless double $\beta$-decay of several nuclei.
\end{abstract}

\section{Introduction}    
The discovery of neutrino oscillations in the atmospheric Super
Kamiokande \cite{SK}, solar SNO \cite{SNO}, reactor KamLAND \cite{Kamland}, accelerator 
K2K \cite{K2K} and
other neutrino oscillation experiments \cite{Cl,Ga,SAGE,SKsol} is one of the major recent discoveries in the
elementary particle physics. All neutrino oscillation data
with the exception of the data of the short baseline LSND experiment
 \cite{LSND} 
are well described
by the three-neutrino mixing \footnote{The LSND result will be checked in the near future 
by the running   MiniBooNE experiment \cite{MiniB}.}
 \be
 \nu_{lL} =
\sum_{i=1}^{3} U_{li} \nu_{iL}\,. 
\label{1} \ee
Here $U$ is PMNS \cite{BP,MNS} mixing matrix, $\nu_{i}$ is the field of neutrino
with mass $m_{i}$ and  $\nu_{lL}$ ($l=e,\mu,\tau$) is 
neutrino field which enter into the standard charged and neutral currents
(so called flavor neutrino field).

From the analysis of the SK atmospheric neutrino data 
for oscillation parameters $|\Delta m^{2}_{32}|$ and 
$\sin^{2}2\theta_{23}$ the following $90 \%~ \rm{CL}$ ranges 
were found
\cite{SK}\footnote{The neutrino mass-squared difference is determined
as follows: $\Delta m^{2}_{ik}=m^{2}_{i}-m^{2}_{k}$. In the case of
normal mass spectrum neutrino masses are labeled in such a way that
$m_{1}<m_{2}<m_{3}$. In this case $\Delta m^{2}_{32}>0$. In the case
of inverted neutrino mass spectrum neutrino masses are labeled as
follows  $m_{3}<m_{1}<m_{2}$. In this case $\Delta m^{2}_{32}<0$.}
\be 
1.9\leq|\Delta m^{2}_{32}|\leq 3\cdot
10^{-3}\,~\rm{eV}^{2};\,~~\sin^{2}2\theta_{23}<0.9
\label{2}
\ee
From analysis of the data of the solar and the reactor KamLAND
experiments for oscillation parameters 
$\Delta m^{2}_{21}$ and $\tan^{2}\theta_{12}$
it was found \cite{SNO}
\be
\Delta m^{2}_{21}=(8^{+0.6}_{-0.4})\cdot
10^{-5}\,~\rm{eV}^{2};\,~~\tan^{2}\theta_{12}=0.45^{+0.09}_{-0.07}.
\label{3}
\ee

From the exclusion plot obtained from the analysis of the data of
the reactor CHOOZ experiment \cite{CHOOZ} for the parameter
$\sin^{2}\theta_{13}$ the following upper bound can be inferred

\be \sin^{2}\theta_{13}< 5\cdot 10^{-2}\,~~(90\%~CL).\label{4}\ee

An information about the absolute values of neutrino masses can be 
obtained from  experiments on the measurement of the high-energy part of the 
$\beta$-spectrum of
tritium and from cosmological data. From the data of the Troitsk \cite{Troitsk} and
Mainz \cite{Mainz} tritium experiments 
it was found
\be 
m_{0}< 2.2 \,~\rm{eV}, 
\label{5}
\ee
where $m_{0}$ is the mass of the lightest neutrino.
From analysis of the cosmological data for the sum of the neutrino masses 
the upper bound in the range
\be 
\sum _{i}m_{i}< (1-2)\,~ 
\rm{eV}. 
\label{6}
\ee
 was obtained (see \cite{Tegmark}).

The nature of the massive neutrinos (Dirac or
Majorana?) is at present unknown. The progress in the
understanding of the origin of the neutrino masses and mixing
strongly depends on the answer to this fundamental question. In particular, the proof
that massive neutrinos $\nu_{i}$ are Majorana particles would  provide a
strong argument in favor of the famous
see-saw mechanism which, apparently, is the most natural mechanism of 
the generation of small neutrino masses.
\section{Neutrinoless double $\beta$-decay}
To reveal the nature of the massive neutrinos it is necessary to study
processes in which the total lepton number $L=L_{e}+L_{\mu}+L_{\tau} $ 
is not conserved. 
The search for neutrinoless double $\beta$-decay
($0\nu\beta\beta$-decay) of some even-even nuclei 
\be 
(A,Z)\to
(A,Z+2)+e^{-}+e^{-}
\label{7}
\ee 
is the most sensitive method of the
investigation of the nature of the massive neutrinos $\nu_{i}$.

If $\nu_{i}(x)$ satisfies
the Majorana condition 
\be \nu_{i}(x)=\nu^{c}_{i}(x)=
C\,\bar\nu^{T}_{i}(x), 
\label{8} 
\ee 
($C$ is the matrix of the
charge conjugation) the process (\ref{7}) is allowed. 
In this case neutrinoless double $\beta$-decay
is a process of the
second order in the Fermi constant $G_{F}$ with virtual neutrinos $\nu_{i}$. The
half-life of the process is given by  the following general expression
 (see reviews \cite{Reviews})
\be \frac{1}{T_{1/2}(A,Z)}=
|m_{ee}|^{2}\,|M(A,Z)|^{2}\,G(E_{0},Z). 
\label{9} 
\ee
Here 
\be m_{ee}=\sum_{i}U^{2}_{ei}\,m_{i}
\label{10}
\ee 
is the
effective Majorana mass, $G(E_{0},Z)$ is  known 
phase-space factor ($E_{0}$ is the energy release) and $M(A,Z)$ is nuclear
matrix element (NME). For light neutrino masses 
NME  do not depend on $m_{i}$.

The results of many experiments on the search for $0 \nu \beta
\beta$ -decay are available at present (see \cite{Elliot}). The most
stringent lower bounds on the half-life of the $0\,\nu \beta\,\beta $-
decay were obtained in the Heidelberg-Moscow
 $^{76} \rm{Ge}$ experiment  \cite{HM} and in the recent CUORICINO $^{130} \rm{Te}$ 
cryogenic experiment \cite{Cuoricino}:
\bea 
T_{1/2}(^{76} \rm{Ge})\geq 1.9 \cdot 10^{25}\, y\,~~(\rm{Heidelberg-Moscow})
\nonumber\\
T_{1/2}(^{130} \rm{Te})\geq 1.8 \cdot 10^{24}\, y~~
(\rm{ CUORICINO}).
\label{11}\eea
Taking into account different calculations of  nuclear matrix
elements, for the effective Majorana mass $|m_{ee}|$ from these
results the following upper bounds can be inferred
\bea |m_{ee}| \leq
(0.3-1.2)\,~\rm{eV}\,~~~(\rm{Heidelberg-Moscow})\nonumber\\
 |m_{ee}| \leq
(0.2-1.1)\,~\rm{eV}\,~~~(\rm{CUORICINO})\label{12} 
\eea
Many new experiments on the search for the neutrinoless double
$\beta$-decay of different nuclei are in preparation at present (see
\cite{Future}). In these future experiments two order of the magnitude
improvement in the sensitivity to $|m_{ee}|$ is expected :\be
|m_{ee}|\simeq \rm{a~~ few}\cdot 10 ^{-2}.
\label{13}
\ee
\section{Effective Majorana mass}
The effective Majorana mass $m_{ee}$ is determined by neutrino
masses $m_{i}$ and elements $U_{ei}$. In the standard
parametrization we have
\be 
U_{e1}=\cos  \theta_{13}~\cos \theta_{12}~e^{i\alpha_{1}};~     
U_{e2}=\cos  \theta_{13}~\sin \theta_{12}~e^{i\alpha_{2}};~
U_{e3}=\sin \theta_{13} ~e^{i\alpha_{3}},
\label{14} 
\ee
where $\alpha_i$ are Majorana CP phases.

From neutrino oscillation data we know the values of neutrino mass-squared differences
$|\Delta m^{2}_{32}|$ and $\Delta m^{2}_{21}$ and the angle
 $\theta_{23}$ (see (\ref{2}) and (\ref{3})). We know also that the angle  
$\theta_{13}$ is small  (see (\ref{4})).
We do not know the character of neutrino mass spectrum (normal
or inverted), the mass of the lightest neutrino $m_{0}$, 
which determine the absolute values of neutrino masses, and Majorana
phases.

Let us consider three standard neutrino mass spectra (see \cite{bbpapers})
\begin{enumerate}
\item
Hierarchy of neutrino masses \be m_{1} \ll m_{2} \ll m_{3}.
\label{15}\ee In this case we have

\bea |m_{ee}|\simeq \left |\sin^{2} \theta_{12} \,
 \sqrt{\Delta m^{2}_{21}} +
\sin^{2} \theta_{13}\, \sqrt{\Delta
m^{2}_{32}}\,e^{2i\alpha_{32}}\right|\nonumber\\ \leq \left
(\sin^{2} \theta_{12} \,
 \sqrt{\Delta m^{2}_{21}} +
\sin^{2} \theta_{13}\, \sqrt{\Delta m^{2}_{32}}\right),
\label{16}
\eea
where $\alpha_{32}=\alpha_{3}-\alpha_{2}$.
Using the values (\ref{2}) and  (\ref{3}) of neutrino oscillation parameters from
(\ref{16}) for effective Majorana mass we find the upper bound 
\be
|m_{ee}|\leq 6.4\cdot 10^{-3}
\label{17}
\ee 
Thus, in the case of the neutrino mass hierarchy 
 the predicted upper bound of $|m_{ee}|$ is significantly smaller than the sensitivity of the most ambitious future experiments.

\item
Inverted hierarchy of neutrino masses 
\be m_{3} \ll m_{1} <m_{2}.
\label{18}
\ee 
In this case we have
\be 
|m_{ee}|\simeq \sqrt{ |\Delta m^{2}_{31}|}\,~ (1-\sin^{2}
2\,\theta_{12}\,\sin^{2}\alpha_{21})^{\frac{1}{2}}\, 
\label{19}
\ee
From this relation we find
\be 
\sqrt{ |\Delta m^{2}_{31}|}\,~ \cos   2\,\theta_{12} \leq
|m_{ee}|\leq \sqrt{ |\Delta m^{2}_{31}|}
\label{20}
\ee
Let us notice that the bounds in this inequality correspond to the case of the CP
conservation in the lepton sector: the upper (lower) bound
corresponds to the case of equal (opposite) CP-parities of $\nu_{2}$
and $\nu_{1}$. 

Thus, in the case of the inverted neutrino mass hierarchy from neutrino oscillation data we can 
predict the upper and lower bounds of the possible values of the effective Majorana mass. 
From (\ref{2}), (\ref{3}) and (\ref{20}) 
we obtain the range
\be 
1.0\cdot 10^{-2}\leq |m_{ee}|\leq
5.5\cdot 10^{-2}~\rm{eV}
\label{21}
\ee 
The values of $|m_{ee}|$ in this
range  apparently will be reached in future
$0\nu\beta\beta$- experiments.
From (\ref{19}) for the parameter $\sin^{2}\alpha_{21}$, which
characterizes CP violation in the case of the Majorana neutrino
mixing, we have
\be \sin^{2}\alpha_{21}=
\frac{1}{\sin^{2}
2\,\theta_{12}}~\left(1-\frac{|m_{ee}|^{2}}{|\Delta
m^{2}_{31}|}\label{22}\right)
\ee
Hence, the observation of the neutrinoless double $\beta$-decay with
$|m_{ee}|$ in the range (\ref{21}) could allow to obtain an
information about Majorana CP phase difference.
\item
Quasi-degenerate neutrino mass spectrum 
\be 
m_{1}<m_{2} <m_{3} ~( m_{3}<m_{1} <m_{2})
;~~m_{0}\gg \sqrt{ |\Delta m^{2}_{32}|}
\label{23}
\ee 
Neglecting
small contribution of $\sin^{2} \theta_{13}$ for the effective
Majorana mass we have in this case
\be 
|m_{ee}|\simeq m_{0}\,~ (1-\sin^{2}
2\,\theta_{12}\,\sin^{2}\alpha_{21})^{\frac{1}{2}}\, 
\label{24}
\ee
In the future tritium experiment KATRIN \cite{Katrin} the sensitivity
$m_{0}\simeq 0. 2$ eV is planned to be reached. 
If the mass of the lightest neutrino
$m_{0}$ will be measured in this experiment for the effective Majorana
mass from (\ref{3}) and (\ref{24}) 
we obtain the range
\be 0.23 ~m_{0}\leq |m_{ee}|\leq m_{0}.
\label{25}
\ee 
On the other side 
if neutrinoless double $\beta$-decay will be observed in future experiments with the value of 
$|m_{ee}|$ which is larger than the upper bound (\ref{21}) for the mass of the lightest neutrino we 
will have the bound 
\be |m_{ee}|\leq m_{0}\leq
4.4~|m_{ee}|.
\label{26}
\ee

\end{enumerate}

\section{Nuclear matrix elements}

The observation of neutrinoless double $\beta$-decay would be a direct
proof that $ \nu_{i}$ are Majorana particles. As we have seen, the
determination of the effective Majorana mass would allow to obtain
an important information about character of the neutrino mass spectrum, mass of the
lightest neutrino and, possibly,  Majorana CP phase difference.

However, from the measurement of the half-life of $0\nu\beta\beta$-decay only the {\em
product} of the effective Majorana mass and nuclear matrix element can be determined.
The calculation of NME is a complicated nuclear problem (see
\cite{Reviews}). Two models are commonly used: Nuclear Shell Model (NSM)
and Quasiparticle Random Phase Approximation (QRPA) with numerous modifications. 
 These two
approaches are based on different physical assumptions. As a result
different calculations of the same NME differ by factor 2-3 or even
more .

 We will discuss now a possible method of a model independent test
 of NME calculations \cite{bbNME}. 
We will use only the general factorization property of matrix elements of the 
$0\nu\beta\beta$ decay. Namely the fact that 
the matrix element of the process is  a product of the effective Majorana neutrino mass, which is determined by neutrino 
masses and mixing, and 
nuclear matrix element, which (for light neutrinos) does not depend on neutrino masses.
From (\ref{9}) we find the following relations between half-lives of the $0\nu\beta\beta$- decay
of different nuclei:
\bea
T_{1/2}(A_{1},Z_{1})=X(A_{1},Z_{1};A_{2},Z_{2})~T_{1/2}(A_{2},Z_{2})=\nonumber\\
X(A_{1},Z_{1};A_{3},Z_{3})~T_{1/2}(A_{3},Z_{3})=....
\label{27}
\eea
Here
\be
X(A_{i},Z_{i};A_{k},Z_{k})=\frac{|M(A_{k},Z_{k})|^{2}\,G(E_{0},Z_{k})}{|M(A_{i},Z_{i})|^{2}\,G(E_{0},Z_{i})}
\label{28}
\ee
The coefficients in relations (\ref{27}) have to be calculated. If $0\nu\beta\beta$-decay of 
{\em different nuclei} will be observed in future experiments and relations (\ref{27}) with coefficients 
$X(A_{i},Z_{i};A_{k},Z_{k})$ calculated in some model $M$ are satisfied than the model  $M$ is compatible 
with data (it is obvious that if 
relations  (\ref{27}) are not satisfied the corresponding model 
must be rejected). This does not mean, however, that the model $M$ allows us to obtain the correct value of $|m_{ee}|$ from experimental data. In fact if nuclear matrix elements,
calculated in the framework of different models $M_{a}$ and  $M_{b}$ are proportional
\be
| M_{M_a}(A,Z)|=\beta ~|M_{M_b}(A,Z)|,
\label{29}
\ee
($\beta$ is a coefficient which does not depend on $(A,Z)$) and relations (\ref{27}) 
are satisfied for the model $M_{a}$ than obviously they are  satisfied also for the model $M_{b}$. The values of 
the effective Majorana mass, which can be determined from experimental data in the framework of these two models,  are connected by the relation
\be
|m_{ee}|_{M_a}=\frac{1}{\beta} ~|m_{ee}|_{M_b}
\label{30}
\ee
and could be quite different.

For the purpose of illustration we will calculate
the 
coefficients in Eq.  (\ref{27}) 
in three different recent models 
of the calculation of nuclear matrix elements of 
$0\nu\beta\beta$-decay: RFSV \cite{RFSV}, CS \cite{CS} and  NSM \cite{NSM}. In the paper  \cite{RFSV}, based on QRPA and renormalized QRPA approaches, 
the values of the parameter of the particle-particle interaction $g_{pp}$, determined from
the measured half-lives of the $2\nu\beta\beta$-decay of corresponding nuclei, were used.
In the QRPA calculation \cite{CS}  the values of the parameters
were determined from the data on the $\beta$-decay of
the nuclei of the interest for double $\beta$-decay transitions.
In the paper \cite{NSM} the results of the latest NSM calculations were given. 
We will consider four different nucleus:
$^{76}\rm{Ge}$, $^{100}\rm{Mo}$,
$^{130}\rm{Te}$ and $^{136}\rm{Xe}$. 
In the Table I we have presented the values of coefficients $X(A_{i},Z_{i};A_{k},Z_{k})$  in the 
case if $0\nu\beta\beta$-decay of $^{76}\rm{Ge}$ and one of the other nucleus is observed.

\begin{center}
                   Table I
\end{center}
\begin{center}
The values of the coefficient  $X(A_{i},Z_{i};A_{k},Z_{k} )$  obtained with NME
calculated in \cite{RFSV} (RFSV), in \cite{CS}  (CS) and in \cite{ NSM} (NSM).

\end{center}
\begin{center}
\begin{tabular}{|cccc|}
\hline

&
$RFSV$
&

$CS$

&

$NSM$

\\

\hline

$X(^{130} \rm{Te}; ^{76} \rm{Ge})$

&

0.38

&

0.13

&

0.24
\\
\hline

$X(^{136} \rm{Xe}; ^{76} \rm{Ge})$

&

0.80

&

0.07

&

0.56
\\
\hline

$X(^{100} \rm{Mo}; ^{76} \rm{Ge})$

&

0.59

&

0.17

&

---
\\
\hline

\end{tabular}
\end{center}
If it would occur that the relation (\ref{27}) with the coefficient, calculated in one of the model considered,
is satisfied, in this case, as we can see from the Table I, other models apparently can be excluded 
(if accuracy of experimental data are better than $\simeq$ 30 \%)

Let us stress that this conclusion depends 
on nuclei for which
neutrinoless double $\beta$-decay is observed. For example, if  $0\nu\beta\beta$-decay of 
$^{130} \rm{Te}$ and $^{100} \rm{Mo}$ will be observed, in this case we have

$$X(^{100} \rm{Mo}; ^{76} \rm{Ge})=1.5 ~(RFSV);~~~ 1.3~ (CS)$$
The difference between the values of these coefficients is only $\simeq$ 10 \%.. Thus, if the relation
(\ref{27}) is satisfied for NME calculated in, say, RFSV model \cite{RFSV},  it will be difficult to exclude the model 
CS \cite{CS}. The values of the effective Majorana mass which can  be obtained from the experimental data in the case of these two models are, however,  quite different:
\be
 |m_{ee}|_{\rm{RFSV}}\simeq 2.5\cdot |m_{ee}|_{\rm{CS}}.
\label{31}
\ee
The observation of $0\nu\beta\beta$-decay of three (or more) nuclei  would be an important tool
in a model independent approach to the determination of the value of the effective Majorana mass  $|m_{ee}|$ which we have discussed here.

\section{Conclusion}
The establishment of the nature of the massive neutrinos $\nu_{i}$ would have a profound importance for the understanding of the origin of small neutrino masses and neutrino mixing.
The investigation of the neutrinoless double $\beta$-decay is the most sensitive method which could allow 
us to reveal the Majorana nature of the massive neutrinos. Today's limit on the effective Majorana mass,
which can be inferred from the study of this process, is $|m_{ee}|\leq (0.3-1.2)$ eV. In several experiments now
in preparation the sensitivity $|m_{ee}|\simeq \rm{a ~few} ~10^{-2}$ eV is planned to be reached. 
If $|m_{ee}|$ is measured, the pattern of the neutrino mass spectrum and, possibly, Majorana CP phase 
could be revealed. For that not only $0\nu\beta\beta$-decay must be observed but also nuclear matrix elements must be known.  Observation of $0\nu\beta\beta$-decay of several nuclei could 
provide a model independent method of testing of NME calculations.
\section{Acknowledgements}

I acknowledge the support of  the Italien Program  ''Rientro dei cervelli''.
It is a pleasure for me to thank Milla Baldo Ceolin for the invitation to the meeting.

\end{document}